\documentclass[aps,prl,reprint,groupedaddress]{revtex4-2}

\usepackage{amsmath}
\usepackage{amsfonts}
\usepackage{graphicx}

\begin{document}

\title{Collective multimode strong coupling in plasmonic nanocavities}

\author{Angus Crookes$^1$}
\author{Ben Yuen$^1$}
\author{Angela Demetriadou$^{1,*}$\thanks{a.demetriadou@bham.ac.uk}}

\affiliation{$^1$School of Physics and Astronomy, University of Birmingham, Edgbaston, Birmingham, B15 2TT, United Kingdom}

\date{\today}

\begin{abstract}
Plasmonic nanocavities enable access to the quantum properties of matter, but are often simplified to single mode models despite their complex multimode structure. Here, we show that off-resonant plasmonic modes in fact play a crucial role in strong coupling, and determine the onset of a novel collective interaction. Our analysis reveals that $n$ strongly coupled plasmonic modes, introduce up to $n(n+1)/2$ oscillation frequencies that depend on their coupling strengths and detunings from the quantum emitter. Furthermore, we identify three distinct regions as the coupling strength increases: (1) single mode, (2) multimode, and (3) collective multimode strong coupling. Our findings enhance the understanding of quantum dynamics in realistic plasmonic environments and demonstrate their potential to achieve ultra-fast energy transfer in light-driven quantum technologies.
\end{abstract}

\maketitle

\section{\label{introduction} Introduction}

Strong coupling between light and matter is essential across various fields, as it connects nanophotonics, materials science, chemistry, and quantum technologies \cite{torma2014strong,flick2018strong,bitton2019quantum}. Strong light-matter coupling enables quantum emitters (QE) to coherently exchange energy with light, creating new polaritonic states that play a critical role in quantum information systems \cite{novotny2012principles}.
To reach the strong coupling regime, quantum emitters (such as molecular dyes, cold atoms and quantum dots) must be placed within regions of intense light, where light-matter interaction rates outpace losses. 
Therefore, plasmonic nanocavities (such as metallic nanostructures, dimers, and nanoparticle on mirror cavities) are ideal for realising strong coupling \cite{chikkaraddy2016single, kleemann2017strong,memmi2017strong,wei2021plasmon,huang2019real,heintz2021few} as localised surface plasmons allow for exceptional subwavelength light confinement and extreme coupling strengths \cite{hugall2018plasmonic, murray2007plasmonic}. In addition, they are straightforward to synthesize, chemically stable, and facilitate the precise positioning and alignment of molecules for reliable and highly reproducible experiments \cite{ojambati2019quantum,liu2024deterministic, chikkaraddy2024single}.

Theoretical descriptions reveal how these new polariton states emerge, and continue to guide experiments towards the generation of new quantum states. Although significant advancements have been made in describing QEs within plasominc nanocavities - accounting for more complex vibrational structures and larger numbers of molecules  \cite{mandal2023theoretical, herrera2020molecular,tokman2023dissipation} - usually the underlying assumption is that they interact with just a single cavity mode.  The single mode approximation originates from the analysis of Fabry-P\'erot and photonic crystal cavities \cite{breuer2002theory,shore1993jaynes,garraway2011dicke,larson2021jaynes}, where modes are spectrally separated relative to their coupling strengths. However, multimode coupling has been shown to play an important role experimentally in some optical cavities \cite{wickenbrock2013collective, sundaresan2015beyond, vaidya2018tunable, mckay2015high}. Plasmonic nanocavities in particular support a dense collection of modes that overlap in frequency, and often all exhibit significant field enhancements \cite{kongsuwan2020plasmonic, elliott2022fingerprinting, bedingfield2023multi}. Although coupling to multiple plasmonic modes has been previously considered \cite{krimer2014route, li2016transformation, medina2021few, sanchez2022few, yang2024electrochemically,crookes2024taming} in general, the underlying origin and impact of the complex  quantum dynamics that emerge in these multimode systems is still not known.  

In this work, we demonstrate that multiple, off-resonant plasmonic modes can significantly dominate the quantum dynamics of QEs in plasmonic nanocavities. In particular, we show that $n$ strongly coupled modes introduce up to $n(n+1)/2$ distinct oscillation frequencies in the QEs excited state population. These frequency components depend on the number of strongly coupled modes, and their respective coupling strengths and detunings from the QE. In fact, we identify three distinct regions defined by the dipole moment: (1) single mode strong coupling, (2) multimode strong coupling, and (3) collective multimode strong coupling. Our results provide a comprehensive understanding of the quantum dynamics in multi-mode environments, and demonstrate how realistic plasmonic nanocavities can be used to achieve ultra-fast energy transfer, for use in light-driven quantum technologies.

\section{RESULTS AND DISCUSSION} 

\begin{figure*}[htbp]
	\includegraphics[width=\textwidth]{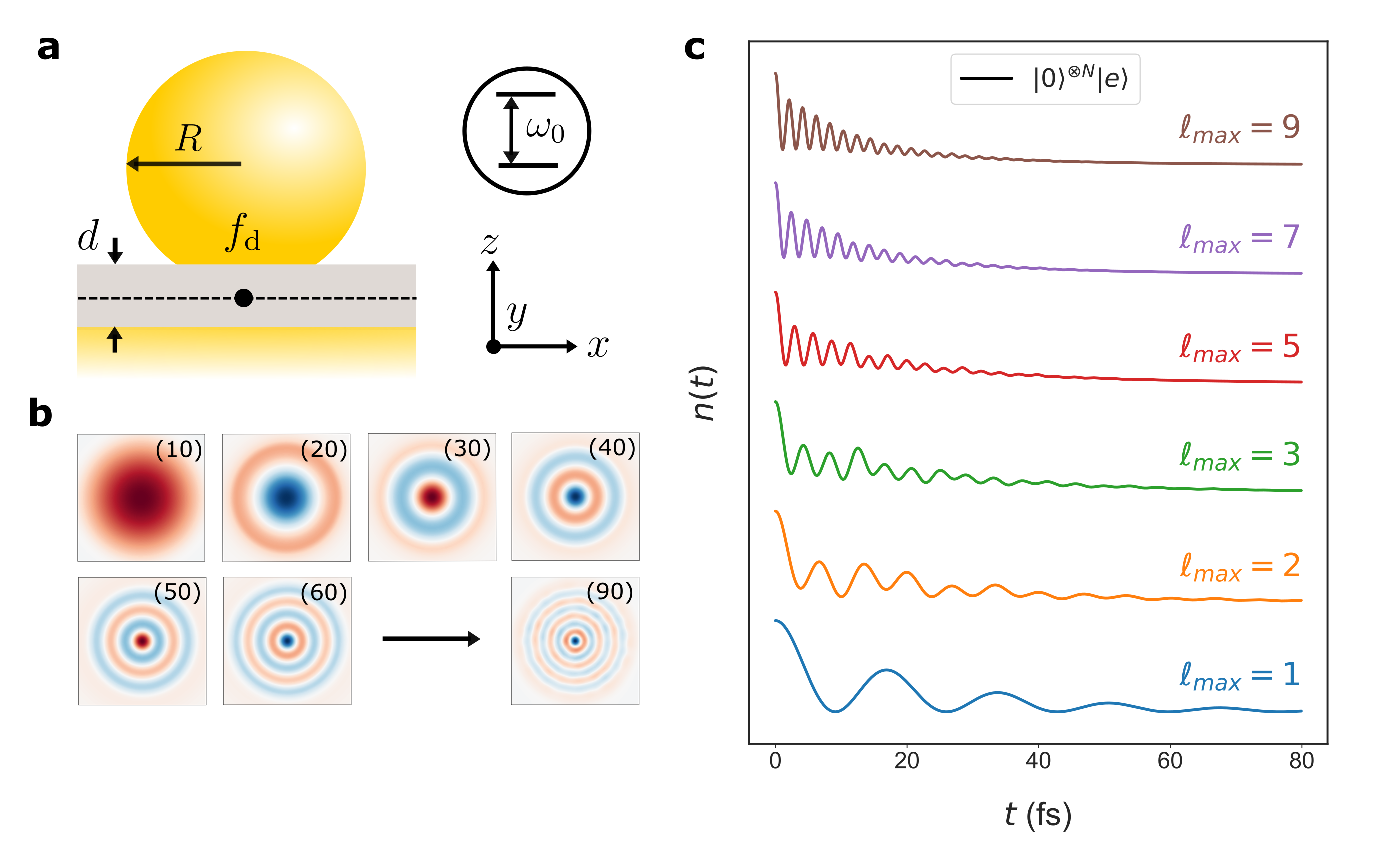}
	\caption{Multimode strong coupling (a) Schematic of a gold nanoparticle on mirror (NPoM) cavity with radius $R = 40$ nm, facet diameter $f_{\text{d}}=16$ nm, gap spacing $d_{\text{gap}}=1$ nm, and gap permittivity $n_{\text{gap}}=2.5$. A single QE is placed in the gap centre at $\mathbf{r}_0=(0,0,0)$. (b) Electric field, $E_{z}(x,y,0)$, of the first nine $\xi=(\ell0)$ quasinormal modes (QNM's). (c) Evolution of the QEs excited state population, $n(t) = |c_0(t)|^2$, including interacting modes up to $\xi = (\ell_{\text{max}}0)$ for various values of the index $\ell_{\text{max}}$.}
	\label{fig:figure_one}
\end{figure*}

The interaction between multiple plasmonic modes and a single quantum emitter (QE) can be described effectively using an open quantum system formalism \cite{bedingfield2023subradiant, crookes2024taming}. In this form, the density operator $\rho(t)$ evolves under: 
\begin{equation}
	\dot{\rho}(t) = -i\left[\mathcal{H},\rho\right] + \sum_{\xi}^n\kappa_{\xi}\left(a_{\xi}\rho a_{\xi}^{\dag}-\frac{1}{2}\{a_{\xi}^{\dag}a_{\xi}, \rho\}\right) 
	\label{master_eqn}
\end{equation}
with Hamiltonian,
\begin{equation}
	\mathcal{H}= \underbrace{\sum_{\xi=1}^n\omega_{\xi} a_{\xi}^{\dag}a_{\xi} + \frac{\omega_0}{2}\sigma_{\text{z}}}_{\mathcal{H}_0} +  \underbrace{\sum_{\xi}^{n} g_{\xi}(\mathbf{r})a_{\xi}^{\dag}\sigma + g^*_{\xi}(\mathbf{r})a_{\xi}\sigma^{\dag}}_{\mathcal{H}_{\text{int}}}
	\label{h_sys}
\end{equation}
where $a_{\xi}^{\dag}$ and $a_{\xi}$ are the creation and annihilation operators for each plasmonic mode $\xi$ with frequency $\omega_{\xi}$ and loss rate $\kappa_{\xi}$, where $n$ plasmonic modes are considered. The raising and lowering operators for a QE with transition frequency $\omega_{0}$ are given by $\sigma^{\dag}$ and $\sigma$ respectively, with the position dependent coupling strength to each mode $\xi$ expressed as $g_{\xi}(\mathbf{r})$ which depends on the QEs position within the plasmonic nanocavity.

One of the most common plasmonic systems that single molecule strong coupling has been realised in is the nanoparticle on mirror (NPoM) cavity \cite{saez2022plexcitonic, lee2022nanoparticle, bedingfield2023subradiant} shown in Figure \ref{fig:figure_one} (a).  Here, we consider an NPoM cavity with nanoparticle radius $R=40$ nm, circular facet size $f_{\text{d}}=16 $ nm, gap size $d = 1$ nm, and gap refractive index $n_{\text{gap}} = 2.5$ - a single QE (two-level system) is placed within the gap at the nanocavity centre i.e. $\mathbf{r}_0=(0,0,0)$. Although we focus on the NPoM cavity, our results apply to any system described by Eq. (\ref{master_eqn}).

The frequencies, loss rates, and interaction strengths of the plasmonic modes - each interacting with the QE through Eq. (\ref{master_eqn}) - are calculated classically using the auxiliary eigenvalue approach \cite{wu2023modal, yan2018rigorous, lalanne2018light}.  These are quasinormal modes (QNMs) with electric and magnetic fields denoted by $\tilde{\mathbf{E}}_{\xi}(\mathbf{r})$ and $\tilde{\mathbf{H}}_{\xi}(\mathbf{r})$ respectively, and have a finite lifetime due to their complex eigenfrequency, which takes the form $\tilde{\omega}_{\xi} = \omega_{\xi} - i\frac{\kappa_{\xi}}{2}$, where $\omega_{\xi}$ represents the resonant frequency and $\kappa_{\xi}$ the mode decay rate. The coupling strength of each QNM $\xi$ to the QE 
is given by:  $g_{\xi}(\mathbf{r}) = \sqrt{\frac{\omega_{\xi}}{\hbar \text{V}_{\xi} }}\boldsymbol{\mu}\cdot\tilde{\mathbf{E}}_{\xi}(\mathbf{r}_j)$ where $\boldsymbol{\mu}$ is the QE dipole moment and $\text{V}_{\xi}=\iiint_{\Omega} \left[\tilde{\mathbf{E}}\cdot\frac{\partial \omega \boldsymbol{\epsilon}}{\partial \omega}\tilde{\mathbf{E}} - \mu_0 \tilde{\mathbf{H}}\cdot\frac{\partial \omega \boldsymbol{\mu}}{\partial \omega}\tilde{\mathbf{H}}\right]dV$ is the QNM normalisation factor. In this system, QNMs are characterised according to their symmetries \cite{kongsuwan2020plasmonic, crookes2024taming} with each given an index $\xi = (\ell m)$, where $\ell\in\mathbb{Z}$ is the number of radial anti-nodes and $-\ell \leq m \leq \ell$ the pairs of azimuthal anti-nodes \cite{kongsuwan2020plasmonic}. The electric field $E_{\xi}^{z}(x,y,0)$ of the $\xi=(\ell0)$ QNMs are shown in Figure. \ref{fig:figure_one} (b) up to $\xi = (90)$. Importantly, the $(\ell0)$ modes are spherical and have their field maximum at the centre of the cavity, facilitating an environment for multi-mode strong coupling. In addition, the QE only interacts with the $(\ell 0)$ modes since the electric field of all other modes vanishes at $\mathbf{r}_0=(0,0,0)$ (see Supplementary Information for more details).

To determine the effect of multiple modes on the quantum dynamics, we first consider a QE in the initial state $c_0(0) = |0\rangle^{\otimes n}|e\rangle$ with dipole moment $\mu = 72~\hat{\textbf{z}} ~ \text{D}$ \cite{ostapenko2010large} and transition frequency resonant with the $(10)$ mode i.e. $\omega_{\text{e}}=\omega_{(10)} = 283$ THz. The quantum dynamics when including
interacting modes up to $n = (\ell_{\text{max}}0)$ are shown in Figure \ref{fig:figure_one} (c) up to  $\ell_{\text{max}} = 9$ and were calculated using diracpy \cite{Diracpy2022}. For a single mode ($\ell_{\text{max}}=1$) the QE exchanges energy with the cavity at a single (Rabi) frequency, which experiences large damping rate due to the high plasmonic losses as expected. However, when two modes interact with the QE ($\ell_{\text{max}}=2$) - with one mode off resonant - the quantum dynamics become increasingly more complex. In this case, the excited state population exhibits multiple oscillations arising from the different coupling strengths and detunings of each mode. In addition, the dominant oscillations in the signal are faster than with a single mode alone. The quantum dynamics continue to exhibit additional multiple oscillation frequencies until, for this system, approximately five modes ($\ell_{\text{max}} \sim 5$) are considered. At this point, a single ultra-fast oscillation frequency dominates the QEs evolution. The energy exchange with the multimode plasmonic nanocavity is almost one order of magnitude faster than when using a single mode approximation. Therefore, from Figure \ref{fig:figure_one} (c) it is obvious that off-resonant modes have a significant impact on the quantum dynamics, and must always be taken into account to accurately describe strong coupling in such systems.

\subsection{QUANTUM OSCILLATIONS IN MULTIMODE STRONG COUPLING} 

The oscillations shown in Figure \ref{fig:figure_one} (c)  arise from the off-resonant modes, but are hard to interpret due to the high plasmonic losses. To overcome this, we initially assume the system evolves in the absence of loss (i.e. $\kappa_{\xi}=0$) such that the evolution is governed by the Schrödinger Equation $i\partial_t|\psi\rangle =\mathcal{H}_{\text{sys}}|\psi\rangle$ where the excited state population of the QE can be determined both numerically and analytically. This helps provide a comprehensive and in-depth understanding of the interactions involved. To derive an equation for the QEs excited state population, we first transform to the interaction picture such that $i\partial_t|\tilde{\psi}\rangle = V_{\text{int}}|\tilde{\psi}\rangle$ where  $V_{\text{int}} = e^{i\mathcal{H}_0t}\mathcal{H}_{\text{int}}e^{-i\mathcal{H}_0t}$ and $|\tilde{\psi}\rangle = e^{i\mathcal{H}_0t}|\psi\rangle$. In fact, due to conservation of the excitation number, we can also express the quantum state as $|\tilde{\psi}\rangle = c_0|0,e\rangle + \sum_j c_{\xi}a_{\xi}^{\dag}|0,g\rangle$ with amplitudes $c_0$ and $c_{\xi}$ respectively. From this, we obtain two coupled equations of motion:
\begin{equation}
	i\partial_tc_0 = \sum_{\xi} g_{\xi}c_{\xi} e^{i\Delta_{\xi}t}
	\label{coupled_eqn_1}
\end{equation}

\begin{equation}
	i\partial_tc_{\xi} = g_{\xi}c_0 e^{-i\Delta_{\xi} t}
	\label{coupled_eqn_2}
\end{equation}
where $\Delta_{\xi} = \omega_0 - \omega_{\xi}$ is the detuning between the QE and mode $\xi$. To solve Eq. (\ref{coupled_eqn_1}) and Eq. (\ref{coupled_eqn_2}) for $c_0(t)$ we take their Laplace Transform, where we define $\tilde{c}(s) = \mathcal{L}\left[c(t)\right]$ and use the identities $\mathcal{L}\left[\partial_t x\right] = s\tilde{x}(s) - x_0(0)$ and $\mathcal{L}\left[f(t)e^{\alpha t}\right] = \tilde{f}(s-\alpha)$. Hence, two new coupled equations are obtained given by $s\tilde c_0(s) - c_0(0) = -i\sum_{\xi} g_{\xi} \tilde c_{\xi} (s-i\Delta_{\xi})$ and $s \tilde c_{\xi}(s) - c_{\xi}(0) = -i g_{\xi} \tilde c(s+i\Delta_{\xi})$ respectively, which we then solve algebraically to give:
\begin{equation}
	\tilde{c}_0(s) = \left[s+ \sum_{\xi}^n \frac{g_{\xi}^2}{(s-i\Delta_{\xi})}\right]^{-1}c_0(0)
	\label{laplace_1}
\end{equation}
where $c_0(0)=1$ is the initial population of the QEs excited state. The key step in solving Eq. (\ref{laplace_1}) is to write the expression in square brackets as a quotient of two polynomial functions:
\begin{equation}
	\left[s+ \sum_j^n \frac{g_j^2}{(s-i\Delta_j)}\right]^{-1}
	= \frac{\text{Q}^{(n)}(s)}{\text{P}^{(n+1)}(s)}
\end{equation}
where
\begin{equation}
	\text{P}^{(n+1)}(s) = s\prod_j^n(s-i\Delta_j) + \sum_{j}g_j^2\prod_{k\neq j}(s-i\Delta_k)
	\label{polynomial_p}
\end{equation}
\begin{equation}
	\text{Q}^{(n)}(s) = \prod_j^n(s-i\Delta_j)
	\label{polynomial_q}
\end{equation}
Importantly, when $\Delta_i\neq\Delta_j$ for all $i\neq j$, then $\text{P}^{(n+1)}(s)$ has $n+1$ distinct and purely imaginary roots $i\lambda_j$ where $\lambda_n < \lambda_{n-1} < ... < \lambda_1 < \lambda_{n+1}$ (see Supplementary Information for more details). This enables us to factorise $\text{P}^{(n+1)}(s)$ and rewrite Eq. (\ref{laplace_1}) in a form that has a simple inverse Laplace Transform: 
\begin{equation}
	\frac{\text{Q}^{(n)}(s)}{\text{P}^{(n+1)}(s)} = \sum_{j=1}^{n+1}\frac{\alpha_j}{s-i\lambda_j}
	\label{inverse_laplace}
\end{equation}
where $\alpha_j = \text{Q}\left(i\lambda_j\right)/\frac{d\text{P}^{(n+1)}}{ds}|_{i\lambda_j}$. Finally, taking the inverse Laplace Transform of Eq. (\ref{inverse_laplace}) now yields an expression for $c_0(t)$, from which we find the QEs excited state population:
\begin{equation}
	\begin{aligned}
		|c_0(t)|^2 &= \sum_{j=1}^{n+1}\alpha_{j}^2 + 2\alpha_{n+1}\sum_{j=1}^n\alpha_j\cos\left(\Omega_j t\right) \\ &\quad + \sum_{j=1}^n\sum_{k\neq j}\alpha_j\alpha_k\cos\left((\Omega_j-\Omega_k)t\right)
	\end{aligned}
	\label{c_0}
\end{equation}
where $\Omega_j = \lambda_j-\lambda_{n+1}$ and therefore, gives the oscillation frequencies present in the quantum dynamics of a multimode system in the limit of zero loss. In fact, Eq. (\ref{c_0}) shows that there are up to $n(n+1)/2$ frequency components in the excited state population of a QE strongly coupled with $n$ modes. More specifically, there are $n$ components corresponding to the mode frequencies $|\Omega_j|$ and $n(n-1)/2$ corresponding to the interference terms $|\Omega_i-\Omega_j|$. 

% add figures
\begin{figure*}[htbp]
	\includegraphics[width=\textwidth]{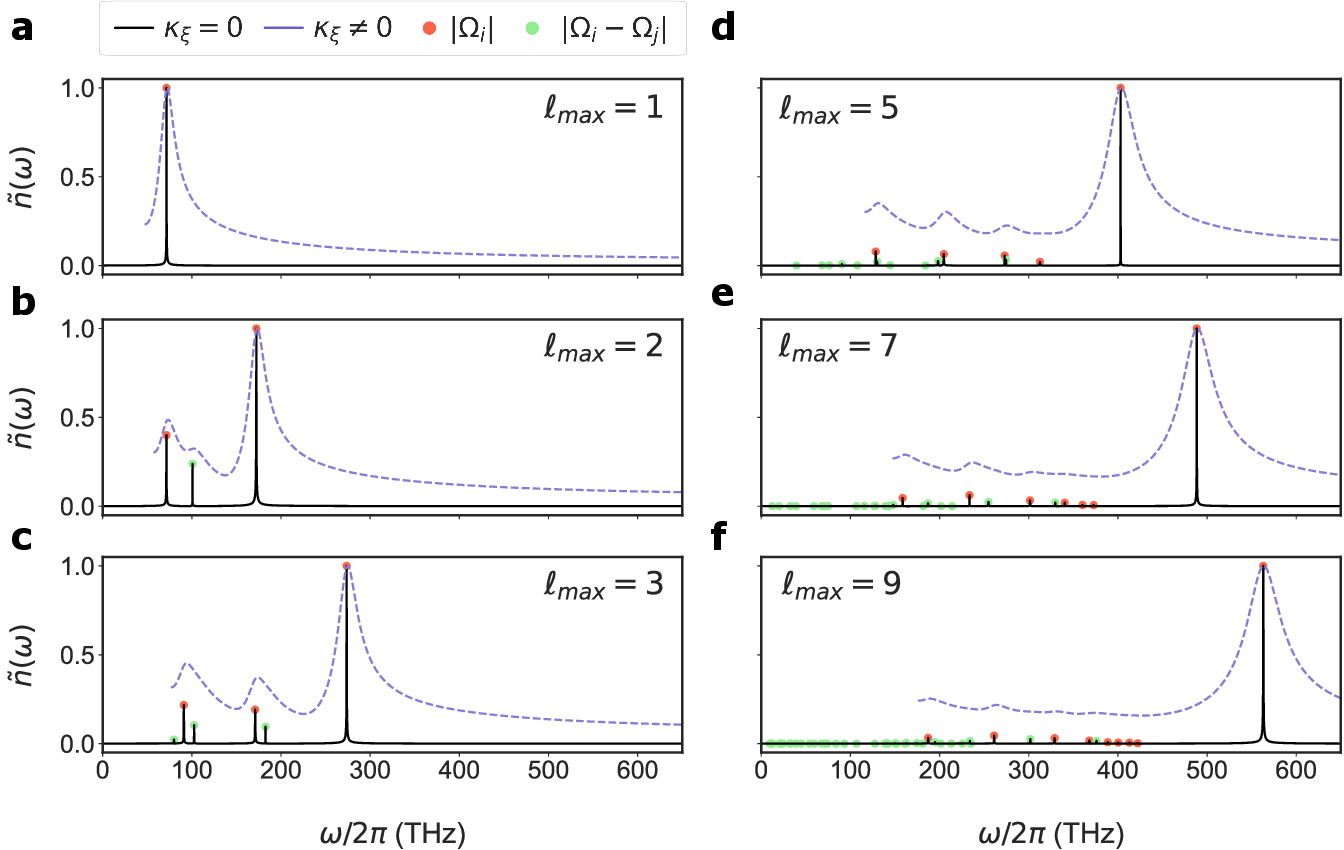}
	\caption{Frequency components in multimode strong coupling. The Fourier Transform $\tilde n(\omega)$ of the QEs excited state population when including interacting modes up to $\xi=(\ell_{\text{max}}0)$. The solid black lines are numerical results without loss (i.e.$\kappa_{\xi}=0$) and the dashed lines with plasmonic loss (i.e. $\kappa_{\xi}\neq 0$). The red and green dots result from  $|\Omega_j|$ and $|\Omega_i-\Omega_j|$ calculated from the roots of Eq. (\ref{polynomial_p}). The sub-figures show the frequencies for when (a) $\ell_{\text{max}}=1$ (b) $\ell_{\text{max}}=2$ (c) $\ell_{\text{max}}=3$ (d) $\ell_{\text{max}}=5$ (e) $\ell_{\text{max}}=7$ (f) $\ell_{\text{max}}=9$ respectively. Note, the spectra with loss are cut to remove the peak at $\omega=0$ which results from the FT of a damped oscillations.}
	\label{fig:figure_two}
\end{figure*}

These oscillations (i.e. $|\Omega_j|$ and $|\Omega_j-\Omega_i|$) obtained in Eq. (\ref{c_0}) are shown in Figure. \ref{fig:figure_two}  in red and green dots, together with the Fourier Transform (FT) in the absence of loss (solid black lines). Adding the plasmonic loss (i.e. $\kappa_{\xi}\neq 0)$ broadens and slightly shifts the spectra (purple dashed line). 
We initially consider one plasmonic mode ($\ell_{\text{max}}=1$) as is often assumed to be the case, and obtain the Rabi oscillation frequency $|\Omega_1|$ as expected. Considering two plasmonic modes ($\ell_{\text{max}}=2$) leads to three distinct peaks, one corresponding to each mode i.e. $|\Omega_1|$ and $|\Omega_2|$ and one interference peak $|\Omega_1-\Omega_2|$ of smaller amplitude. Here, all three peaks have a high intensity and contribute to the QEs evolution, but the off-resonant mode is the most significant. This is not always the case, and as we see later depends on the number and properties of the strongly coupled modes (i.e. their coupling strengths and detunings).
As we increase the number of modes strongly coupled with the QE we continue to observe up to $n(n+1)/2$ oscillations in the quantum dynamics. For a large number of modes (i.e. $\ell_{\text{max}}\geq 5$) the quantum dynamics are completely dominated by a single ultra-fast frequency component $|\Omega_n|$ with only a small contribution from other components. It is extremely important to note, that although we associate the frequency component $|\Omega_j|$ with the mode $j$, each peak still depends on all the other strongly coupled modes in the cavity. In fact, the $|\Omega_j|$, mode $j$ association is lost the more strongly coupled modes are considered, as each mode gradually shifts from being a distinct peak to contributing to the single `supermode' frequency component $|\Omega_n|$. 
Note that this ultra-fast oscillation $|\Omega_n|$ emerges from the combined effect of all modes, and does not occur with the same amplitude and frequency if only some of the modes are considered (see Supplementary Information). Therefore, this is a collective phenomenon and necessitates a full description of the plasmonic environment.

\begin{figure*}
	\includegraphics[width=\textwidth]{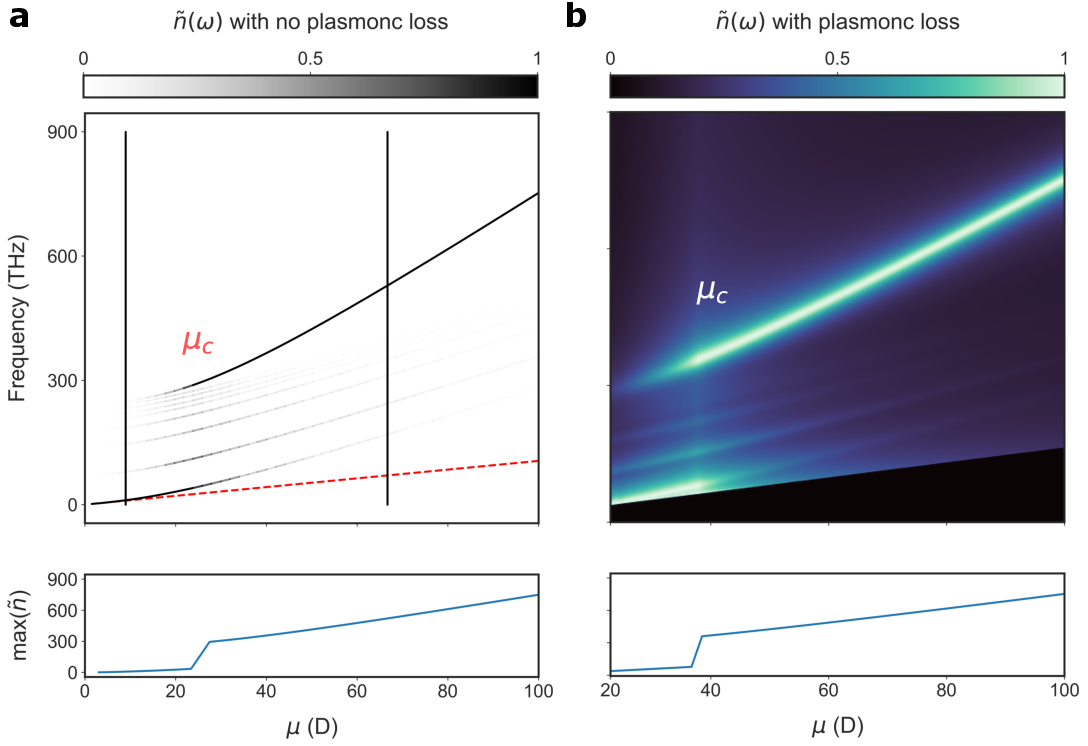}
	\caption{Transition from single to multimode strong coupling. Fourier Transform of the QEs excited state population as a function of dipole moment $\mu$. The QE interacts with modes up to $\xi=(90)$ with (a) assuming no loss i.e. $\kappa_{\xi}=0$ and (b) including loss i.e. $\kappa_{\xi}\neq0$. Three regions, dictated by the dipole moment, govern the QEs evolution ($I$) single mode strong coupling ($II$) multimode strong coupling and ($III$) collective multimode strong coupling. The single mode approximation is shown in the dashed red dashed line. The critical dipole moment is also shown, and represents the onset of the ultra-fast collective strong coupling. }
	\label{fig:figure_three}
\end{figure*}

\subsection{CRITICAL TRANSITION TO COLLECTIVE STRONG COUPLING}

The amplitudes of the oscillations at frequencies $|\Omega_j|$ and $|\Omega_j-\Omega_i|$ depend on the magnitude of the coupling strengths $g_j$ compared to the detunings $\Delta_j$. To determine the impact of the coupling strength on the oscillations reported above, we change the dipole moment $\mu$ (which is an overall scale factor of the coupling strengths) and identify three distinct regions. In Figure \ref{fig:figure_three} (a) and (b) the FT of the quantum dynamics is shown for a system with and without loss respectively for many dipole moments. For small dipole moments such that $g_i < |\Delta_i - \Delta_{i + 1}|$ for each mode (region $I$) the dynamics are determined by a single resonant mode and leads to a single (Rabi) oscillation. This applies to the majority of molecules commonly utilized in plasmonics, such as Cy5 and methylene blue, indicating that for QEs with small $\mu$ single mode strong coupling is reached.  However, when plasmonic losses are present in this system, damping prevents single mode strong coupling for the same dipole moments. For intermediate dipole moments (i.e. quantum dots) such that $|\Delta_i - \Delta_{i+1}|^2 < |g_i|^2 << | \Delta_{i+1} | | \Delta_i - \Delta_{i+1} |$ (region $II$), then multi-mode strong coupling occurs. In this region, the off-resonant modes produce multiple distinct frequency peaks  $|\Omega_j|$, while the interference components $|\Omega_i-\Omega_j|$ have a small amplitude and are hard to observe (further discussion is included in the Supplementary Information for $\mu = 30$ D). Finally, for large dipole moments such that  $|g_i|^2 >> | \Delta_{i+1} | | \Delta_i - \Delta_{i+1} |$ (region $III$) multimode strong coupling is still significant, and in fact the quantum dynamics are now governed by a single ultra-fast `supermode' oscillation $|\Omega_n$|. This `supermode' oscillation results in energy exchange over an order of magnitude faster than single mode strong coupling alone. In region $II$ the supermode peak $|\Omega_n|$ can have the largest amplitude at some dipole moments, but in region $III$ it completely dominates the quantum dynamics. The quantum dynamics in Figure \ref{fig:figure_one} and \ref{fig:figure_two} for $\mu = 72$ D are within region $III$ and therefore show the collective interaction with multiple off-resonant modes.

Crucially, a transition occurs at a critical dipole moment  ($\mu_{\text{c}}$), where the dominant frequency component switches from $|\Omega_1|$ to $|\Omega_n|$ as shown in Figure  \ref{fig:figure_three} (bottom panels). At the critical dipole moment, the maximum oscillation frequency in a lossless system increases from $34$ to $294$ THz at $\mu_c \sim 27$ D and from $75$ to $360$ THz at $\mu_c \sim 38$ D in an equivalent system with loss. The critical dipole moment occurs at higher values with loss because dissipation suppresses the amplitude of the off-resonant peaks and so requires larger coupling to reach criticality. As the dipole moment increases, the distinct resonances begin to overlap and interfere. In the limit of large coupling, the oscillation amplitudes $\alpha_i$ switch from behaving like $[1+\lambda_i \sum_{j=1}^n 1/(\lambda_i-\Delta_j)]^{-1}$ to $[1-\sum_{j\neq k} g_k^2 / ((\lambda_i-\Delta_j)(\lambda_i - \Delta_k))]^{-1}$ (see Supplementary Information for the derivation). The former has the most dominant oscillations for $\lambda_1$ i.e. the component oscillating at $\Omega_1$, while the latter has the most dominant oscillations for $\lambda_n$, i.e. the component oscillating at $\Omega_n$. This occurs because the interactions with each mode interfere constructively for the amplitude $\alpha_n$ when $\mu>\mu_c$, whilst there is always some destructive interference for the other amplitudes. Therefore, past the critical dipole moment (which occurs at some point within between two regions) energy exchange with the plasmonic nanocavity becomes ultra-fast, and represents a collective multi-mode strong coupling regime. 

In plasmonic systems, multiple plasmonic modes cannot be neglected; the single-mode approximation holds only for small dipole moments, which are often unachievable in practical setups due to the inherent plasmonic losses. As the coupling strength increases, the interaction transitions to a fully collective regime characterized by a dominant ultra-fast oscillation governing the quantum dynamics. This behaviour highlights a key advantage of quantum dots (QDs) in quantum plasmonics: with dipole moments surpassing 
$70$ D (as seen in InGaN/GaN QDs) as they can reach the collective multimode strong coupling regime. As a result, QDs can produce ultra-fast oscillations nearly two orders of magnitude faster than those achievable with dye molecules, and an order of magnitude faster than single mode approximations. In addition, collective multimode strong coupling can also be attained in plasmonic systems with further reduced mode volumes, enabling critical coupling even at lower dipole moments. These findings present a novel pathway for ultrafast coupling in multimode systems, with significant potential for applications in quantum information and sensing.

\section{Conclusions} 

In conclusion, we have demonstrated the crucial importance of off-resonant plasmonic modes strongly coupled to QEs. We show that $n$ strongly coupled plasmonic modes introduce up to $n(n+1)/2$ oscillation frequencies in the excited state population of a single QE going far beyond the single mode approximation. Additionally, we identify three distinct coupling regimes based on increasing dipole moment: single-mode strong coupling, multi-mode strong coupling, and collective multimode strong coupling. These oscillations are influenced by the number of modes, their coupling strengths, and detunings. These results deepen our understanding of quantum dynamics in multimode environments and demonstrate the potential for plasmonic nanocavities to achieve ultra-fast energy transfer in light-driven quantum technologies.
\newline

\noindent \textbf{Funding:} AD gratefully acknowledges support from the Royal Society University Research Fellowship URF\textbackslash R1\textbackslash 180097 and URF\textbackslash R\textbackslash 231024, Royal Society Research grants RGS \textbackslash R1\textbackslash 211093, funding from ESPRC grants  EP/Y008774/1 and EP/X012689/1. AD, BY acknowledge support from Royal Society Research Fellows Enhancement Award RGF \textbackslash EA\textbackslash 181038, and AD, AC acknowledge funding from EPSRC for the CDT in Topological Design EP/S02297X/1.
\newline

\noindent \textbf{Author Contributions:} All the authors have accepted
responsibility for the entire content of this submitted
manuscript and approved submission.
\newline

\noindent \textbf{Conflict of Interest:} Authors state no conflict of interest.
\newline 

\noindent \textbf{Data Availability Statement:} The datasets generated and/or analysed during the current study are available from the corresponding author upon reasonable request.

\bibliographystyle{apsrev4-2}

%\bibliography{arxiv_bib}

%

\end{document}